\documentclass[12pt]{article} 

\usepackage{epsfig} 
\usepackage{amsfonts} 
\usepackage{amssymb}
\usepackage{amsmath}  
\usepackage{amsbsy}  
\usepackage{wasysym}

\catcode`\@=11 
\@addtoreset{equation}{section} 
\def\theequation{\thesection.\arabic{equation}} 
 
\@twosidetrue 

\catcode`\@=11  
\def\section{\@startsection{section}{1}{\z@}{3.5ex plus 1ex minus 
.2ex}{2.3ex plus .2ex}{\large\bf}}

\def\thesection{\arabic{section}} 
\def\thesubsection{\arabic{section}.\arabic{subsection}} 
\def\thesubsubsection{\arabic{section}.\arabic{subsection}.\arabic{subsubsection}} 
 
\def\appendix{\setcounter{section}{0} 
 \def\thesection{\Alph{section}} 
 \def\theequation{\Alph{section}.\arabic{equation}} 
\def\thesubsection{\Alph{section}.\arabic{subsection}} 
\def\thesubsubsection{\Alph{section}.\arabic{subsection}.\arabic{subsubsection}} 
 
\def\section{\@startsection{section}{1}{\z@}{3.5ex plus 1ex minus 
   .2ex}{2.3ex plus .2ex}{\large\bf}} } 
 
\newcount\minutes 
\newcount\scratch 
 
\def\timestamp{%
\scratch=\time 
\divide\scratch by 60 
\edef\hours{\the\scratch} 
\multiply\scratch by 60 
\minutes=\time 
\advance\minutes by -\scratch 
---$\,$\hours:\null 
\ifnum\minutes< 10 0\fi 
\the\minutes} 

\def\sla#1{\ifmmode%
\setbox0=\hbox{$#1$}%
\setbox1=\hbox to\wd0{\hss$/$\hss}\else%
\setbox0=\hbox{#1}%
\setbox1=\hbox to\wd0{\hss/\hss}\fi%
#1\hskip-\wd0\box1 } 

\begin{document} 
\begin{titlepage} 
\nopagebreak 
{\flushright{ 
        \begin{minipage}{5cm}
         KA--TP--35-2007  \\            
         SFB/CPP-07-94 \\ 
        \end{minipage}        } 
 
} 
\vfill 
\begin{center} 
{\LARGE \bf 
 \baselineskip 0.5cm 
QCD corrections to hadronic WWZ production with leptonic decays
} 
\vskip 0.5cm  
{\large   
V. Hankele and D. Zeppenfeld 
}   
\vskip .2cm
{\it Institut f\"ur Theoretische Physik, 
        Universit\"at Karlsruhe, P.O.Box 6980, 76128 Karlsruhe, Germany}
 \vskip 1.3cm     
\end{center} 
 
\nopagebreak 
\begin{abstract}
Multi-lepton signatures appear in many new physics searches at the Large
Hadron Collider. 
We here consider WWZ production with subsequent leptonic decay of the 
three vector bosons as a SM source of multi-lepton events. We have calculated 
the next-to-leading order QCD corrections for the full 
$p p\to 6$~lepton production cross sections in hadronic collisions.
Results have been implemented 
in the form of a flexible parton-level Monte-Carlo program which allows 
to calculate the QCD corrections for arbitrary distributions and acceptance 
cuts.
\end{abstract} 
\vfill 
\hfill 
\vfill 

\end{titlepage} 
\newpage               
%
%
\section{Introduction}
\label{sec:intro}

With the start of the CERN Large Hadron Collider (LHC) a trove of new
data is anticipated, which can be used to search for new physics
and also to further probe the Standard Model (SM).
For the interpretation of these data, precise predictions for both
the desired signal processes and backgrounds are needed, and this 
necessitates the calculation of next-to-leading order (NLO) QCD 
corrections. In order to determine cross sections for non-trivial 
acceptance cuts, it is most useful to cast these NLO calculations 
into fully flexible Monte-Carlo programs which
can calculate cross sections as well as distributions.

WWZ production with subsequent leptonic decay of the vector
bosons is a background to supersymmetric processes with several
leptons in the final state. In addition, it is an excellent probe of the 
quartic electroweak $W^+ W^-\gamma \gamma$, $W^+ W^- Z \gamma$ and 
$W^+ W^- Z Z$ couplings. 
Constraints for some of these couplings are already available from the Large
Electron Positron collider (LEP) at CERN~\cite{Alcaraz:2006mx} and analogous 
measurements have been suggested for proton antiproton collisions at the 
Fermilab Tevatron collider~\cite{Han:1995es}. However, the LHC will be able 
to improve these measurements considerably and therefore more accurate 
predictions are needed~\cite{Eboli:2000ad}. As we will see, QCD corrections 
increase the WWZ cross section by more than 70\% and, thus, any
quantitative measurement of quartic couplings will have to take QCD 
corrections into account. 
This sizable enhancement is not surprising in view of the similarity with diboson
production~\cite{Ohnemus:1995gb,Campbell:1999ah}. Furthermore QCD
corrections to ZZZ production at the LHC have already been
calculated and increase the LO cross section by about 50\%
\cite{Lazopoulos:2007ix}.

We here present first results on the NLO QCD corrections to the full $2\to 6$ 
process  $p p \to \nu_e \, e^+ \, \mu^- \, \bar{\nu}_\mu \, 
\tau^- \, \tau^+$ (or any other combination of leptons from three distinct
families). All resonant and non-resonant matrix elements as well as
the spin correlations of the final state leptons are included in our 
calculation. For simplicity, we neglect any identical lepton effects
which might appear when using the results for final states with leptons from
one or two generations only. The calculation is performed with the 
Catani-Seymour subtraction algorithm~\cite{Catani:1996vz} and uses 
the virtual amplitudes derived in Refs.~\cite{Oleari:2003tc,Jager:2006zc}. 
Further details of the calculation and the checks we
performed will be given in section 2. First determinations
of cross sections, distributions and $K$-factors are presented in section 3. 
 
%
%
\section{The Calculation}
\label{sec:Calc}

We have considered in our calculation the full set of Feynman graphs
for the process $p p \to \nu_{\ell_1} \, \ell_1^+ \, \ell_2^- \,
\bar{\nu}_{\ell_2} \, \ell_3^- \, \ell_3^+$ up to order $\alpha_s\;
\alpha^6$. This includes the  Higgs contribution and all off-shell
diagrams.
However, interference terms due to identical particles in the
final state have been neglected. Including such effects at LO we find
that this is an excellent approximation: LO cross sections change by
less than 0.1\% when interference terms are integrated over the Breit
Wigner peaks. Interference terms show strong cancellations between
contributions below and above the Breit Wigner peaks, but even their
absolute values contribute at the few percent level only, which is 
below the scale variation of our final NLO cross sections.
We also neglect any fermion mass
effects. In particular, the Higgs Yukawa coupling to tau leptons is set
to zero. Because of the large number of Feynman diagrams we use the
helicity method of Ref.~\cite{Hagiwara:1985yu} for the evaluation of the
matrix elements.

\begin{figure}[tbp]
\includegraphics[scale = 1]{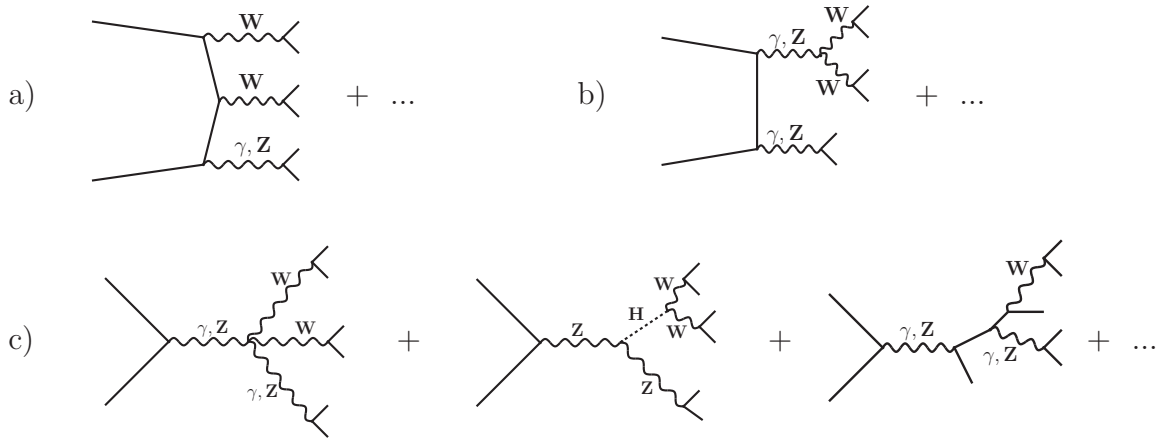}

\caption[]{\label{fig:LO}
{\sl Some representative tree-level Feynman diagrams of the process $p p
  \to 4 \, \ell + 2 \, \nu$. They show the three
  different topologies appearing in this calculation.}
}
\end{figure}

Some representative tree-level Feynman diagrams are given in
Fig.~\ref{fig:LO}. The 181 tree-level graphs can be grouped into 
three distinct topologies. In
Fig.~\ref{fig:LO}a three vector bosons with subsequent decay are
emitted from the quark line. The polarization vectors of these vector 
bosons are the decay currents describing the respective decay leptons, and
these decay currents appear in many different Feynman graphs. In order to
speed up the calculation, they are determined numerically at the beginning 
of the evaluation of the matrix elements for a given phase space point 
and reused wherever they appear.
In Fig.~\ref{fig:LO}b two vector bosons are attached to a
quark line and then decay into two or four leptons. All the Feynman graphs
for a 4-lepton decay can again be combined to an effective polarization 
vector. For all subprocesses like $u \bar{u} \to 4 \, \ell + 2 \, \nu$ or 
$\bar{d} d \to 4 \, \ell + 2 \, \nu$ these polarization vectors are
the same for  one specific phase space point and they do not depend on
the quark polarization.  Furthermore they appear in various Feynman
diagrams. In our code these polarization vectors are therefore
calculated once per  phase space point, stored and reused wherever possible.
The last topology is the one where only one vector boson is attached to
the quark line. The polarization vector corresponding to the ``decay'' of 
this virtual vector boson can be calculated once per phase
space point, stored and reused.
The method of precalculating effective polarization vectors renders our 
code for the Born processes about 4 times faster than a direct evaluation
with {\tt MadGraph}~\cite{Stelzer:1994ta}. 

The full NLO cross section consists of real
emission contributions, virtual contributions and a finite remainder
from the factorization of collinear singularities into the parton distribution
functions (pdfs). 
The real emission contributions have to be integrated
over an (m+1)-particle phase space with $m=6$ leptons, 
and the other two contributions only involve an 
m-particle phase space. 
\begin{align}
\sigma^{NLO} = \int_{m+1}  d\sigma^R + \int_m d\sigma^V +
\int_m d\sigma^C
\end{align}
 The virtual and the real emission contribution are
separately infrared divergent and only their sum gives a well
defined finite result. We have used the dipole subtraction algorithm
proposed by Catani and Seymour \cite{Catani:1996vz} in order to handle
these divergencies in our Monte-Carlo program.

The real emission matrix elements can be obtained from the Born level
matrix elements by either attaching a gluon to a quark line or having a
gluon in the initial state which then splits into a quark anti-quark
pair. The method of precalculating the effective polarization vectors for 
leptonic decays, described above for the tree-level diagrams, has also 
been used for the more complicated real emission diagrams. Here an increase
in computational speed by about a factor of 12 is reached.

The ``virtual'' contribution consists of the square of tree-level diagrams
and the interference between tree-level diagrams and the virtual
one-loop diagrams.
In the calculation of the one-loop diagrams, three different types of
contributions corresponding to the three topologies in Fig.~\ref{fig:LO}
appear. For the simplest topology, with one vector boson attached to the
quark line, only vertex corrections appear. In this case the one-loop
contribution is proportional to the Born matrix element. The second
topology, Fig.~\ref{fig:LO}b, leads to propagator corrections, vertex
corrections and boxes. In the calculation, the sum of these
corrections to one specific tree-level subamplitude is grouped 
together and will be called a boxline contribution in the following. 
The most challenging topology, Fig.~\ref{fig:LO}a, leads to quark propagator 
corrections, vertex corrections, boxes and pentagons. As in the previous 
case the sum of all these corrections to one specific tree level Feynman graph is 
grouped together and called pentline contribution in the following.

The boxline contribution has the same structure as in the vector 
boson fusion process $qq\to Vqq$, which was considered in 
Ref.~\cite{Oleari:2003tc}.  Similarly the pentline contribution
is obtained by crossing from the results of Ref.~\cite{Jager:2006zc}, 
where QCD corrections to the vector boson fusion processes $qq\to qqVV$ were
determined. The singular contributions of all of these
three types of virtual contributions are proportional to the Born matrix
element and the complete virtual one-loop contribution for the three topologies
is given by
\begin{align}
M_V = \tilde{M}_V + \ \frac{\alpha_S}{4 \pi} \ C_F \ \left( \frac{4
      \pi \mu^2}{s} \right)^\epsilon  \ \Gamma{(1 + \epsilon)} \ \left[
    -\frac{2}{\epsilon^2} - \frac{3}{\epsilon} - 8 + \frac{4 \ \pi^2}{3}
  \right] \ M_B,
\end{align}
where $M_B$ denotes the full Born amplitude, $s$ is the square of the
partonic center of mass energy, i.e. it corresponds to the invariant mass
of the 6-lepton system, and $\tilde{M}_V$ is the finite part of the virtual
boxline and pentline amplitudes, which are obtained by crossing and analytic 
continuation from the results of Refs.~\cite{Oleari:2003tc,Jager:2006zc}. 
In the calculation of the boxline routine the usual
Passarino-Veltman tensor decomposition~\cite{Passarino:1978jh} is
applied, whereas in the pentline routine the method proposed by Denner
and Dittmaier~\cite{Denner:2002ii} has been used for the tensor
coefficients of pentagon diagrams.

Since the pentagon routines are quite time consuming we have employed a
method suggested in Ref.~\cite{Jager:2006zc} to reduce the magnitude of the 
true pentagon contribution.
It is possible to shift the polarization vectors or respectively the
decay currents of the vector bosons by terms proportional to their momenta,
\begin{align}
\epsilon_{V}^\mu  = x_{V} \, q_{V}^\mu +
\tilde{\epsilon}_{V}^\mu. \label{eq:eps}
\end{align}
The pentagons
contracted with the momenta instead of the polarization vector 
(terms proportional to $x_V)$ can then
be expressed in terms of boxes via Ward identities for the loop integrals
and the magnitude of the remaining pentagon contribution is
reduced to a numerically less challenging level.
In practice we choose 
\begin{align}
\tilde{\epsilon}_{W^\pm} \cdot (q_{W^+} + q_{W^-}) = 0 \label{eq:eps2}
\end{align}
which means that the shifted polarization vectors have zero time
component in the center of mass system of the W-pair. 

We have checked our calculation at numerous levels.
All the matrix elements for the LO process and the real emission part
have been compared with {\tt MadGraph}~\cite{Stelzer:1994ta} output.
For individual matrix elements we find agreement at the $10^{-15}$ level. In
addition, the total LO cross section has been checked against
{\tt MadEvent}~\cite{Stelzer:1994ta} and {\tt
  HELAC}~\cite{Cafarella:2007pc}. 
For the real emission part the
LO process $p p \to 4 \, \ell + 2 \, \nu + j$ has been compared
against {\tt MadEvent}. Results agree within the statistical accuracy of the
Monte-Carlo runs (0.5\% for {\tt MadEvent} and 0.2\% for {\tt HELAC}).

The finite collinear terms have been checked by exploiting the fact that
they are generic for all Drell-Yan type processes and hence should be
exactly the same for WW production. We have independently programmed NLO
QCD corrections for WW production and compared our
results with {\tt MCFM}~\cite{Campbell:1999ah}. We 
find full agreement for different factorization and renormalization 
scales. As a further check we have taken advantage of the fact that we can
integrate terms proportional to the Born matrix element either together
with the real emission part, by integration over the (m+1)-particle  phase
space, or together with the virtual part, by integrating over the m-particle
phase space. Resulting cross sections are independent of this choice of
procedure.

For the virtual contributions Ward identity tests have been 
implemented as a flag for numerically unstable evaluation of tensor
coefficients. In the case of pentagon diagrams, for example, the pentline 
contribution can be expressed in terms of boxes if one of the external 
polarization vectors is replaced by its momentum. We discard the pentline
contribution at phase space points 
where the two ways of calculating these terms disagree by more than 10\%.
The resultant loss in cross section is negligible: extrapolating its size
we estimate the resultant relative error on the NLO cross section to be
below $10^{-4}$.

Many of the critical elements of our calculation also appear in the related
process $p p \to$ ZZZ for which NLO QCD corrections have recently been
presented in Ref.~\cite{Lazopoulos:2007ix}, albeit without further decay
of the leptons and neglecting any Higgs contributions. 
As a last and very powerful test we have modified our code to describe
ZZZ instead of WWZ production and checked against the results of 
Ref.~\cite{Lazopoulos:2007ix}. Using the same pdfs
and $\alpha_s \, (m_Z) = 0.119$ at NLO we find complete agreement for
all of the given renormalization and factorization scales.
%
%
\section{Results}
\label{sec:res}
The complete next-to-leading order calculation has been implemented in 
the framework of a
fully flexible Monte-Carlo program, {\tt VBFNLO}, which then has been used
to determine the QCD corrections to WWZ production at the LHC.
For the generation of our results we set the electroweak parameters to the
following values:
\begin{align}
&m_W = 80.419 \ \mathrm{GeV}  &&m_Z = 91.188 \ \mathrm{GeV} \nonumber\\
&G_F = 1.16639 \cdot 10^{-5} \ \mathrm{GeV}^{-2} 
&&m_H = 120 \ \mathrm{GeV}. \label{eq:ew}
\end{align}
The other electroweak parameters, $\alpha^{-1} = 132.507$ and
$\sin^2{(\theta_W)} = 0.22225$, are calculated in the program by using LO
electroweak relations.
The default value for the renormalization and factorization
scale is the invariant WWZ mass, which is given by
\begin{align} \label{eq:WWZmass}
\mu_F = \mu_R = \sqrt{(p_{\ell_1} + p_{\ell_2} + p_{\ell_3} + p_{\ell_4} +
  p_{\nu_1} + p_{\nu_2})^2}.
\end{align}
For scale variation studies of the total cross section, we also 
consider fixed renormalization and factorization scales, which we set as
$\mu_F = \mu_R = \xi\cdot m_Z$.
We use the CTEQ6M parton distribution with $\alpha_s(m_Z) = 0.118$ at NLO and
CTEQ6L1 for the LO calculation ~\cite{Pumplin:2002vw}. All fermions are
treated as massless and we do not consider contributions involving
bottom and top quarks. The CKM matrix is approximated by a unit matrix
throughout, which is appropriate when neglecting fermion mass effects in
a neutral Drell-Yan type process.

In order to keep most of the cross section we apply only minimal
cuts to the final state leptons. This means that besides a cut on 
the minimal transverse momentum and the maximal
rapidity of the charged leptons we only require that the invariant
dilepton mass, $m_{\ell \ell}$, of any combination of the charged leptons is 
larger than 15 GeV, thus steering clear of the virtual photon singularity 
in $\gamma^* \to \ell^+ \, \ell^-$ at low $m_{\ell \ell}$. Specifically,
we require  
\begin{equation}
p_{T_\ell} > 10 \ \mathrm{GeV} \qquad\qquad |y_\ell| < 2.5 \qquad\qquad
m_{\ell \ell} > 15 \ \mathrm{GeV} \label{eq:cuts}
\end{equation}
in all subsequent figures. All results given here have been calculated for the
process $p p \to \nu_e \, e^+ \, \mu^- \, \bar{\nu}_\mu \, \tau^- \,
\tau^+$, i.e. interference terms due to identical particles have been 
neglected. In order to obtain cross sections for the phenomenologically 
more interesting case of final states with four electrons and/or muons, 
we have multiplied these results by a combinatorial
factor of 8 in all figures.

\begin{figure}[tbp]
\centerline{
\includegraphics[scale=1]{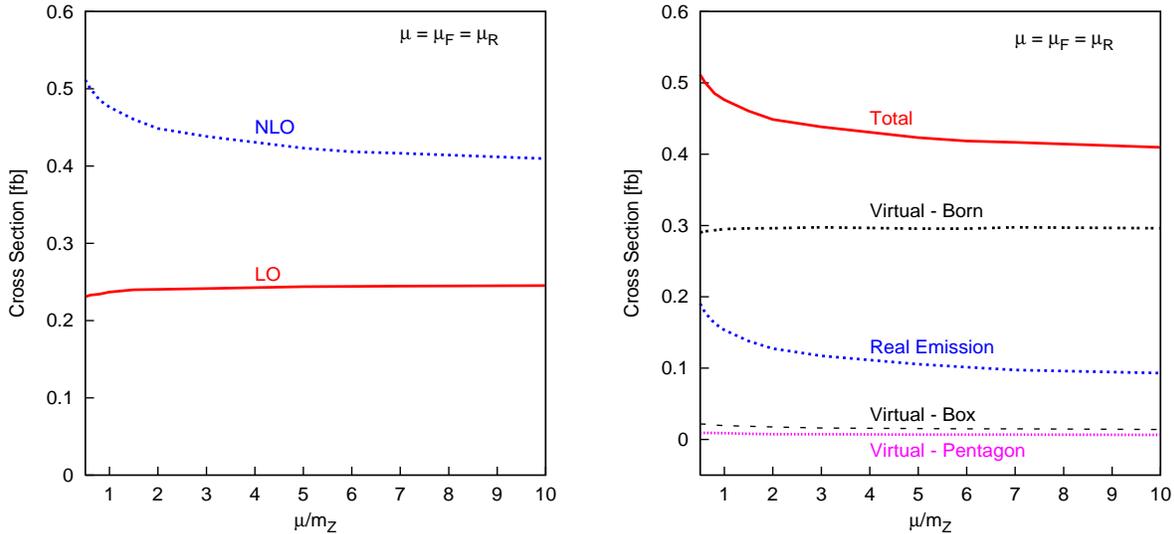}
}

\caption[]{\label{fig:Scale}
{\it Left:} {\sl Scale dependence of the total cross section for 
  $p p \to WWZ \to 4 \, \ell+\sla{p}_T$ at LO and
  NLO for $m_H = 120 \ \mathrm{GeV}$ and the cuts of Eq.~(\ref{eq:cuts}).
  The factorization and renormalization scales are taken at a fixed value 
  which is varied in the range from $0.5 \cdot m_Z$ to $10 \cdot m_Z$.} 
{\it Right:} {\sl Same as in the left panel but for the different NLO
  contributions.}
}
\end{figure}

In the left panel of Fig.~\ref{fig:Scale} the scale dependence of the
cross section at LO and NLO is shown. The LO scale dependence
severely underestimates the uncertainties of the LO cross section 
due to the fact that no $\alpha_S$ appears in the LO calculation and that
the pdfs are determined in a Feynman-$x$ range of small factorization scale
dependence.
The right panel of Fig.~\ref{fig:Scale} shows the scale dependence of the 
different contributions to the NLO cross section. The
virtual contribution is split up into one part proportional to the Born
matrix element, one part with boxline contributions and a third part
which includes only the true pentline contribution after the reduction
described in equations~(\ref{eq:eps}), (\ref{eq:eps2}). This pentagon
contribution amounts to 1 - 2\% of the total NLO cross section.
The real emission contribution in the plot includes the additional 
finite terms from the factorization of collinear singularities into the pdfs.
The scale dependence of the NLO cross section is mainly due this real
emission contribution.

\begin{figure}[tbp]
\centerline{
\includegraphics[scale=1]{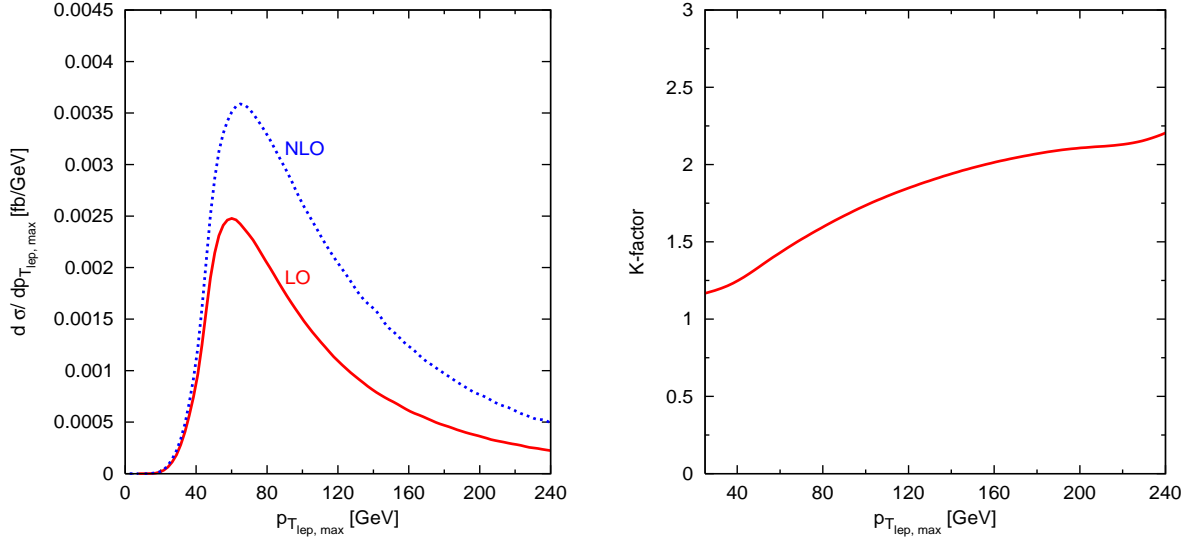}
}

\caption[]{\label{fig:pT}
{\it Left:} {\sl Transverse momentum distribution of the highest-$p_T$ charged
  lepton for $m_H = 120 \ \mathrm{GeV}$, $\mu_F = \mu_R = m_{WWZ}$ as
  given in Eq.~(\ref{eq:WWZmass}) and the cuts
  given in Eq.~(\ref{eq:cuts}) at LO and NLO.}
{\it Right:} {\sl Differential $K$-factor as defined in
  Eq.~(\ref{eq:kfactor}) for the two distributions in the left panel.}
}
\end{figure}

\begin{figure}[tbp]
\centerline{
\includegraphics[scale=1]{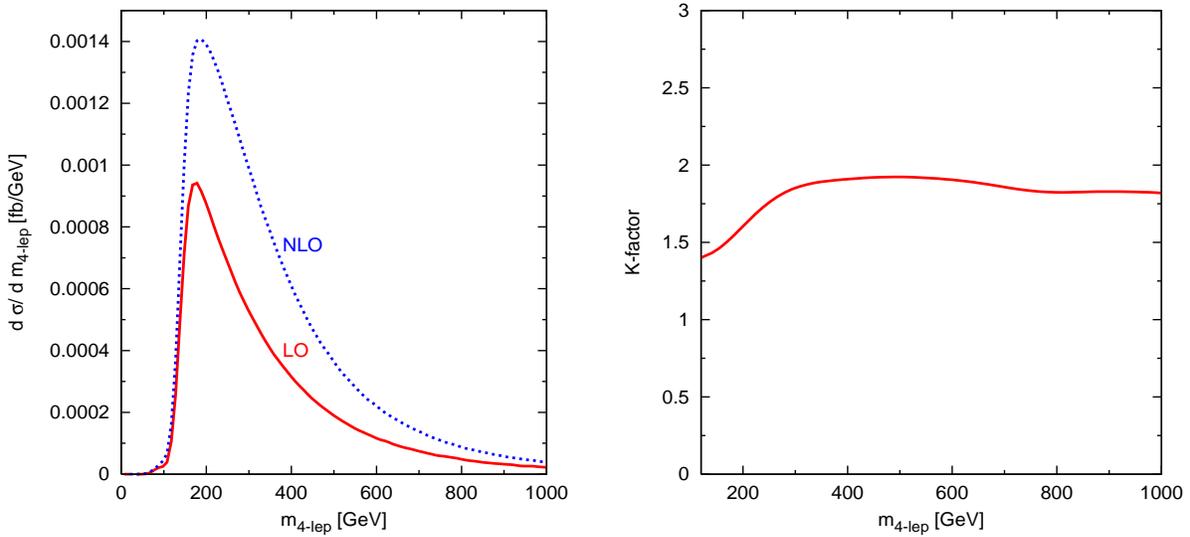}
}

\caption[]{\label{fig:4lmass}
{\it Left:} {\sl Invariant mass distribution of the four charged
  leptons for $m_H = 120 \  \mathrm{GeV}$, $\mu_F = \mu_R = m_{WWZ}$ 
  and the cuts given in Eq.~(\ref{eq:cuts}) at LO
  and NLO.}
{\it Right:} {\sl Differential $K$-factor as defined in
  Eq.~(\ref{eq:kfactor}) for the two distributions in the left panel.}
}
\end{figure}

The $K$-factor, defined as the ratio of the NLO cross section over the LO
cross section strongly depends on the factorization and renormalization
scale. It is shown in Fig.~\ref{fig:Scale} for a fixed scale choice,
$\mu_R=\mu_F=\xi\cdot m_Z$. One finds variations of the $K$-factor 
between 1.7 (for large scales) and 2.2 (for small scales). 
A qualitatively similar scale dependence is found if instead of the
Z-mass the invariant 6-lepton mass as given in Eq.~(\ref{eq:WWZmass}) is
taken as reference scale choice. The mean value of this dynamical
reference scale is substantially larger than the Z-mass and the K-factor for
$\mu_F = \mu_R = m_{WWZ}$ is 1.74. Since the WWZ invariant mass is the scale
of the Drell-Yan type subdiagrams in Fig.~1, it may be considered the most
natural scale choice in the present calculation. We leave a more thorough
investigation of scale dependence of the NLO cross section to a future 
publication~\cite{wwz_long}, however.

The size of the NLO QCD corrections shows a marked phase space dependence.
Two examples are shown in the remaining figures.
In the left panel of Fig.~\ref{fig:pT} the transverse momentum
distribution of the highest-$p_T$ charged lepton is shown for the cuts given in
Eq.~(\ref{eq:cuts}). The right panel shows the differential $K$-factor,
defined as
\begin{align}\label{eq:kfactor}
K = \frac{d \sigma^{NLO}/ dx}{d \sigma^{LO} /dx}.
\end{align}
The $K$-factor increases with $p_T$ by almost a factor 2, indicating 
that a simple multiplication with a constant overall $K$-factor 
would seriously change the shape of lepton $p_T$ distributions. 
A somewhat more benign behavior is found for the 4-lepton invariant 
mass distribution which is shown in Fig.~\ref{fig:4lmass}. Here
the differential $K$-factor varies between 1.4 and 1.9 and most of the
rise is in the threshold region. For 4-lepton invariant  masses above
250~GeV a constant $K$-factor would be an adequate approximation for the
$m_{4 \ell}$ distribution. 


%
\section{Conclusions}
\label{sec:concl}

WWZ production with subsequent leptonic decays is an important
source of multi-lepton events and of great interest for the
measurement of quartic electroweak couplings at the LHC. 
In this letter we have presented first results on the NLO QCD
corrections to the process $p p \to \nu_{\ell_1} \, \ell_1^+ \,
\ell^-_2 \, \bar{\nu}_{\ell_2} \, \ell_3^- \, \ell_3^+$.
For
the WWZ invariant mass as renormalization and factorization scale, the overall
$K$-factor is about 1.7 and therefore has to be taken into account in
any analysis involving WWZ production.

The scale dependence of the NLO cross section is substantially larger
than the variation observed for the LO results. This can be quantified by
increasing and lowering the renormalization and factorization scale by a
factor of 2 around $3 \cdot m_Z$ as central value. At LO the scale
dependence is very small with a variation of less than $\pm$ 1.5\%,
whereas at NLO variations of $\pm$ 5\% appear. The NLO uncertainties
which are indicated by the scale dependence are, thus, typical for a NLO
QCD prediction, while the LO case must be considered as
anomalous. Indeed, the WWZ cross section provides another example where
the scale variation of a LO cross section does not give a good
estimate for the corrections due to higher order effects.

The NLO QCD corrections do not only change the normalization of the
total WWZ cross section, they also lead to substantial shape changes of
distributions: the differential $K$-factors show a sizable variation
over phase space. Reliable modeling of the lepton distributions arising
from WWZ production at the LHC thus requires the inclusion of NLO QCD
corrections. These corrections are now available in the form of a
flexible parton level Monte Carlo program, which will be incorporated
into the publicly available {\tt VBFNLO} package~\cite{vbfnlo} in the
near future. 

%
%
\section*{Acknowledgments}
We would like to thank Frank Petriello for his help in comparing the
ZZZ production cross sections, Malgorzata Worek for the comparison
with {\tt HELAC} and Carlo Oleari for many useful discussions.
This research was supported by the Deutsche
Forschungsgemeinschaft via the Sonderforschungsbereich/Transregio 
SFB/TR-9 ``Computational Particle Physics'' and the Gradu\-iertenkolleg
``High Energy Physics and Particle Astrophysics''.
%
%


\end{document}